\documentstyle[graphicx]{mn2e}

\def\gs{\mathrel{\raise0.35ex\hbox{$\scriptstyle >$}\kern-0.6em 
\lower0.40ex\hbox{{$\scriptstyle \sim$}}}}
\def\ls{\mathrel{\raise0.35ex\hbox{$\scriptstyle <$}\kern-0.6em 
\lower0.40ex\hbox{{$\scriptstyle \sim$}}}}

\title[The dwarf galaxy population in Abell~2218]{The dwarf galaxy 
population in Abell~2218}
\author[Michael.~ B.~ Pracy et al.]{
\parbox[t]{\textwidth}{
       Michael B.~Pracy$^1$, Roberto De~Propris$^{1,2}$, Simon P.~Driver$^2$, 
       Warrick J.~Couch$^1$, and Paul E.J. Nulsen$^3$}  
\\
\vspace*{6pt}\\
$^1$School of Physics, University of New South Wales, Sydney NSW 2052, 
Australia \\
$^2$Research School of Astronomy \& Astrophysics, Australian National University,
 Weston, ACT 2611 \\
$^3$School of Engineering Physics, University of Wollongong, NSW 2522}

\date{Received 0000; Accepted 0000}

\pagerange{\pageref{firstpage}--\pageref{lastpage}}
\pubyear{2001}

\begin{document}

\maketitle

\label{firstpage}
             
\begin{abstract}
We present results from a deep photometric study of the rich galaxy
cluster Abell 2218 ($z=0.18$) based on archival HST WFPC2 F606W
images. These have been used to derive the luminosity function to
extremely faint limits ($M_{\rm F606W} \approx -13.2$ mag, 
$\mu_{0}\approx 24.7$\,mag\,arcsec$^{-2}$) over a wide field of
view ($1.3\; h^{-2}\;$Mpc$^{2}$). We find the faint-end slope of the
luminosity function to vary with environment within the cluster, going
from $\alpha = -1.23\pm 0.13$ within the projected central core of the
cluster ($100 < r < 300\,h^{-1}$\,kpc) to $\alpha = -1.49\pm 0.06$ outside
this radius ($300 < r < 750\,h^{-1}$\,kpc). We infer that the core is
'dwarf depleted', and further quantify this by studying the ratio of
`dwarf' to `giant' galaxies and its dependency as a function of
cluster-centric radius and local galaxy density. We find that this
ratio varies strongly with both quantities, and that the dwarf galaxy
population in A2218 has a more extended distribution than the giant
galaxy population.

\end{abstract}


\section{Introduction}
The galaxy luminosity function (LF) -- the number density of galaxies
per unit luminosity interval -- is a fundamental descriptor of the
galaxy population and, as such, contains important information on the
formation and evolution of galaxies (Bingelli, Sandage \& Tammann 1988, 
Benson et al. 2003). At low redshift, the LF will contain the
combined imprints of the galaxy initial mass function (Press \& 
Schechter 1974, Schechter 1976), together with the effects of any
subsequent evolutionary processes which modified this distribution (e.g.,
hierarchical merging; White \& Rees 1978). Determinations of the LF in
different environments and at a variety of redshifts provide the only
direct hope of disentangling these environmental and evolutionary effects.

In this context, rich clusters are fundamental testing grounds,
representing the densest environments in which galaxies reside and,
within the hierarchical clustering framework, the ultimate examples of
where build-up through merging has taken place. Here galaxies at {\it
all} luminosities, both bright and faint, may bear evidence of
environmental effects: the `giant' ($L>L^*$) galaxies at the bright
end of the LF are the strongest candidates for having been built up
through successive mergers and accretion (e.g. Kauffmann, White \&
Guiderdoni 1993).  Equally, the sub-luminous ($L<L^*$) `dwarf' galaxy
population may be a victim of, and therefore depleted by, 
these merging and cannibalisation processes. In contrast, galaxy `harassment'
may boost  the dwarf galaxy population through whittling down more luminous
cluster galaxies as they undergo high speed encounters (Moore et al. 1996).
Additionally, both giant and dwarf galaxies in the core regions of clusters
are likely to be susceptible to such processes as ram-pressure stripping
(Gunn \& Gott 1972) and cluster tidal effects (Byrd \& Valtonen 1990, 
Bekki et al. 2001), which severely modify their star forming activity and
hence their luminosity (see also Bower \& Balogh 2003, Moore 2003). 

This potential for the rich cluster LF to be used as an evolutionary `probe' 
has led to numerous efforts to measure it. In recent times, the faint-end 
slope of the LF -- described by the parameter, $\alpha$, in the 
Schechter (1976) function representation -- has been of particular focus,
with numerous claims that it is significantly steeper in clusters than in
the general field  (De Propris \& Pritchet 1998, Trentham \& Tully
2002).

Comparison of cluster LF results is, however, complicated by a number
of factors: the differing magnitude ranges over which the slope is
calculated, the different methods of field galaxy subtraction employed by
different authors, and the different physical cluster cross-sections 
covered (Driver \& De Propris 2003). A further complication is that
some observations suggest the numbers of dwarf galaxies varies with
environment -- what Phillipps et al. (1998) referred to as a ``dwarf 
population--density'' relation. This manifests itself as the densest
regions in clusters having the lowest dwarf galaxy fractions, 
the ratio of dwarf to giant galaxies increasing with decreasing giant galaxy 
density, and the dwarf populations in clusters being more spatially extended 
than the giants (Smith et al. 1997, Driver et al. 1998a, Driver et al.
2003). This is also compounded by the likelihood that there exist two dwarf
populations, such as seen in Virgo, whereby the dwarf ellipticals are
centrally concentrated (e.g., Binggeli, Sandage \& Tammann 1985; Conselice
et al. 2003) and the dwarf irregulars (and/or dwarf low surface brightness
galaxies) lie predominantly in the halo (Sabatini et al. 2003). 

In this paper we present a case study of the well known rich
cluster Abell~(A)2218 at $z=0.18$, deriving its LF to very faint limits
and over a broad range of environment from deep, wide-field imaging 
obtained with the {\it Hubble Space Telescope} (HST). Famous for its
spectacular strong gravitational lensing of background galaxies (Kneib 
et al. 1996), A2218 is an Abell richness class 4 cluster with a Bautz Morgan 
type of II, whose centre is dominated by a large, low surface brightness cD
galaxy with an envelope extending over more than 180h$^{-1}$\,kpc 
(Pello-Descayre et al. 1988). It also has a massive, deep potential 
well, with a central velocity dispersion of 1370\,km\,s$^{-1}$ (Le Borgne
et al. 1992), and is a strong X-ray emitter 
[$L_{X}$(0.5--4.4\,keV)$=3.3\times 10^{44}\,h_{100}^{-2}$\,erg\,s$^{-1}$;
Jones et al. 1993]. Moreover, a strong Sunyaev-Zeldovich decrement is
observed in its direction (Birkinshaw \& Hughes 1994). As well as providing
an extreme case in the context of observing environmental effects, A2218 is
also a propitious choice in that its high richness makes it particularly
amenable to reliable `background' removal using statistical
methods (Driver et al. 1998b).

The layout of this paper is as follows: In section 2 we describe the
HST imaging of A2218, and the procedures followed in the detection
and photometry of objects, star/galaxy separation, and the extraction
of the cluster population through the careful subtraction of the 
background population. For the latter, considerable attention is paid
to the numerous effects that contribute uncertainties to this process, 
in particular the effect of gravitational lensing. In section 3 we present
the luminosity function, subdivided into inner (projected core) and outer
(halo) components. We then determine the dwarf-to-giant ratio versus giant
galaxy density and investigate the radial profiles of the giant, dwarf and
ultra-dwarf galaxies (defined below). We summarise and discuss our results
in section 4. Throughout we adopt a  ${\Omega}_{\rm M}=0.3$, ${\Omega}_
{\Lambda} =0.7$ and $H_{0}=100 {\rm km s^{-1} Mpc^{-1}}$ cosmology; this
puts A2218 at a distance modulus of $39.01$\,mag (inclusive of a
$k$-correction), with 1\,arcmin projecting to 128 co-moving kpc at this
distance. 

\section{Data Analysis}

Our study is based on archival HST
\footnote{Based on observations made with the NASA/ESA Hubble Space
Telescope, obtained from the data archive at the Space Telescope
Science Institute. STScI is operated by the Association of
Universities for Research in Astronomy, Inc. under NASA contact NAS
5-26555.}  Wide Field Planetary Camera 2 (WFPC2) images of A2218, obtained as
part of a program to study weak gravitational lensing (PI: Squires,
PID:7343) over a much larger region of this cluster than that covered in
the original strong-lensing study (Kneib et al. 1995). These data comprise a
mosaic of 22 pointings in the F606W passband. Each pointing involved a 
total exposure time of 8,400\,s. The location of these pointings on the sky
is shown in Figure 1; the mosaic provides coverage of a roughly square 
field with a total area of $\approx 82\; {\rm arcmin}^{2}$. The 
resolution, depth and coverage of the data are unique and particularly 
well suited to investigating A2218's galaxy population down to very
faint limits and out to a radius of $\sim 0.75\, h^{-1}$Mpc from its 
centre. 

\begin{figure}
{\includegraphics[width=8cm]{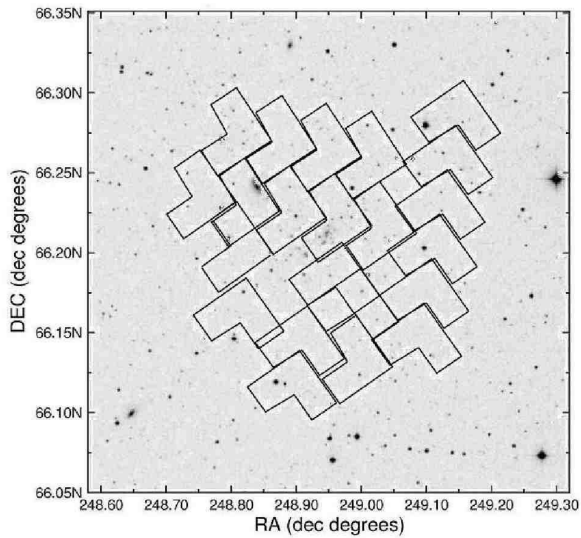}}
\caption{The HST WFPC2 mosaic used for this study, showing the arrangement
and location of the 22 pointings on the sky. A total exposure time of 8,400\,s
was obtained at each pointing. This is overlayed on top of a digital sky 
survey image of the A2218 field.}
\end{figure}

The HST data were obtained from the HST archive with the standard
procedures of bias removal and flat-field division having been applied.  Each
pointing consists of 12 dithered exposures, each of 700\,s duration. In
order to remove cosmic rays, the individual frames were drizzled
together using the {\sc drizzle} task in the IRAF dither package, 
following the procedure of Koekemoer et al. (2002). The final drizzled
image had a pixel size of 0.0498\,arcsec. Only the portions of each 
pointing with the full 8,400\,s of exposure were used in the
analysis, resulting in a final overall field-of-view of 
0.0228\,sq.\,degrees or 1.34\,Mpc$^{2}$ at the cluster redshift.  
The photometric zero point adopted for our data was that published 
for the WFPC2 by Holtzman et al. (1995), which places the instrumental
F606W magnitudes onto the VEGA system. The quoted accuracy of this 
zeropoint is 2\%.

\subsection{Object detection and photometry}

Objects were detected and photometered using the {\sc SExtractor}
package of Bertin \& Arnouts (1996).  Based on previous experience in 
using {\sc SExtractor} with WFPC2 images, a detection criterion of 20 
connected pixels 1.5$\sigma$ above the sky RMS was used. 
Figure 2 shows a typical image 
obtained with one of the WFPC2 CCDs (WF3), with the objects 
detected by {\sc SExtractor} circled. The magnitude {\sc best\_mag} 
was used as our measure of total galaxy magnitude (hereafter denoted 
simply as $F606W$). This adopts a Kron 
magnitude as the default value, except for crowded regions where an 
extrapolated isophotal magnitude is used.  Note that radius of the Kron (1980) 
extraction aperture is set to 2.5 times $R_k$ where $R_k$ is the 
first moment of the image distribution.  A total of 8038 objects were 
detected with $F606W\le 25.8$\,mag (equivalent to $M_{\rm 
F606W} \le -13.2$ mag), the magnitude at which the number of detections
turned over.

\begin{figure}
{\includegraphics[width=8cm]{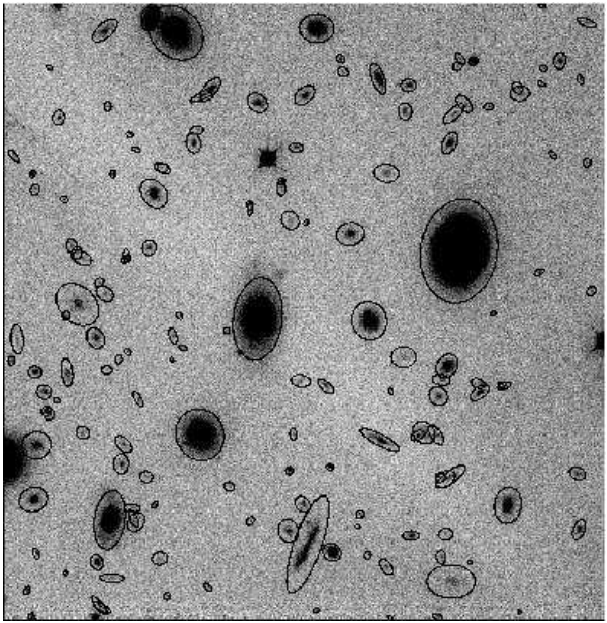}}
\caption{A typical image obtained with one of the WFPC2 CCDs (WF3)
 with {\sc SExtractor} detections circled. The image is $80^{\prime \prime}$ 
 ($0.17\,h^{-1}$\,Mpc) on a side.}
\end{figure}
 
The {\sc SExtractor} parameter {\sc class\_star} was used as
the basis for separating stars from galaxies, supplemented by
visual inspection in the small number of cases where the software
was equivocal in its classification. 

A broad overview of the range and depth of the objects detected by
{\sc SExtractor} and in particular the population of galaxies 
included in this study, is shown in Figure 3. Here we plot 
the mean central surface brightness, $\mu_{\rm app}$ (measured 
in a circular aperture of area equivalent to the 20-pixel minimum 
required for object 
detection), of all our detected objects as a function of their
apparent total magnitude. This reveals three distinct populations
of object: cosmic rays, which form the tight upper-most diagonal
locus, stars, which form the slightly broader central locus, and
galaxies, which form the lower-most broad wedge of objects.
The stars and galaxies are clearly distinguishable down to 
$F606W\sim 24$. Also of note is the surface-brightness `edge' to
the galaxy population, with no objects seen fainter than
$\mu_{\rm app}\sim 24.5$\,mag\,arcsec$^2$. This is consistent with
the $1.5\sigma$ isophotal detection limit used in {\sc SExtractor}, 
which corresponds to $\mu\approx 24.7$\,mag\,arcsec$^{-2}$. Given the 
way the galaxy wedge broadens with increasing magnitude and surface 
brightness, it is clear that our sampling of the galaxy
population is surface-brightness limited at magnitudes fainter
than $F606W\sim 23$. However, 
as we discuss in the following section, this has little impact
on our ability to accurately quantify the cluster galaxy
counts at these faint apparent magnitudes. 

\begin{figure}
\vspace*{0.5cm}
{\includegraphics[width=5.7cm,angle=270]{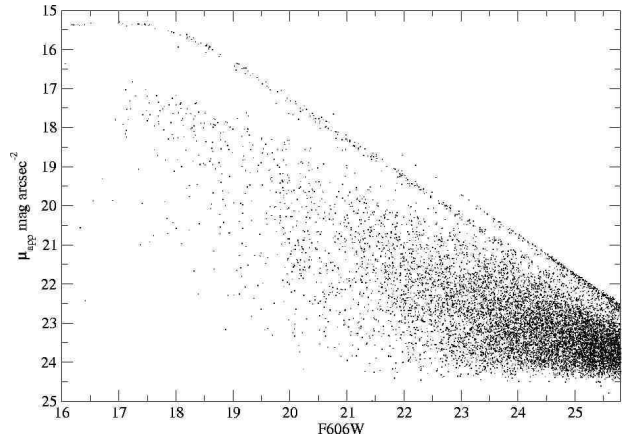}}
\caption{The aperture surface brightness of all our detected objects, 
plotted as a function of apparent magnitude. The measured 
objects, represented by the {\it dots}, are clearly seen to define three 
distinct loci within this diagram: cosmic-rays (upper-most locus), 
stars (central locus), and galaxies (lower wedge).  Stars and
galaxies are easily separated down to a magnitude of $F606W\sim 24.5$.}
\end{figure}

\subsection[]{Field galaxy removal}

In the absence of spectroscopy, the removal of the foreground and
background 'field' galaxy populations must be done statistically.
Here we follow the strategy adopted by Driver et al. (2003) for 
a similar HST-based study of the rich cluster Abell~868 ($z=0.15$). This
involved estimating the field galaxy counts from the deepest available 
HST/WFPC2 data taken in the F606W band. This comprised images of
the Hubble Deep Field (HDF) North (Williams et al. 1996), the HDF-South 
(Casertano et al.  2000), and 10 other fields with total exposure 
times between 10,000--20,000\,s (and hence of comparable depth to
our A2218 cluster mosaic). Object detection and photometry was
performed on these images using {\sc SExtractor}, with an 
identical set of parameters to those used on our A2218 data.

The galaxy number counts derived from these data are shown in Figure 4, 
where we have plotted the combined counts from the two HDF fields and
those from the 10 other reference fields separately. To within the
uncertainties, we see that there is no significant difference in the 
galaxy counts between the `deep' (HDF) and `shallow' reference fields. 
This is important, since by including the much deeper HDF data (surface
brightness limit of $\mu_{\rm app}\sim 26.0$\,mag\,arcsec$^2$ cf. limit
of $\mu_{\rm app}\sim 24.5$\,mag\,arcsec$^2$ for our `shallow' reference
fields and A2218 -- see Fig. 3) in our field galaxy count estimates, 
we want to be certain that this does not lead to an over-estimation 
through the inclusion of an additional population of faint, low surface
brightness galaxies unseen in our cluster images. The agreement seen 
between the two different sets of number counts shown in Figure 4 would
indicate this is not the case. Hence we can be
confident that our selection limits are identical both inside and
outside the cluster, and the excess galaxy counts we observe after
field subtraction is due to having detected a {\it bona fide} population 
of cluster galaxies rather than one that is spurious.

\begin{figure}
{\includegraphics[width=5.7cm,angle=270]{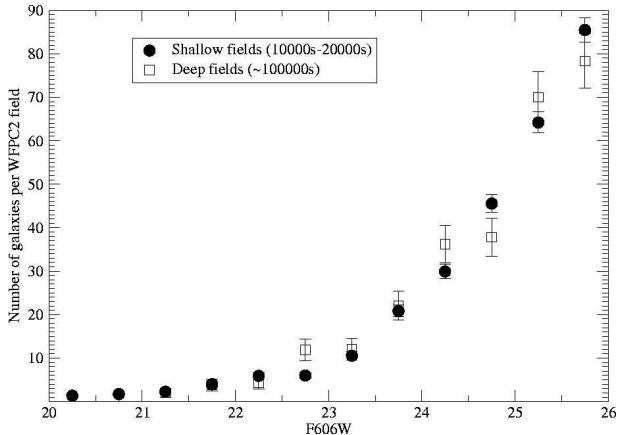}}
\caption{The combined HDF-North and HDF-South galaxy counts ({\it open 
squares}) and the combined shallower reference counts ({\it closed 
circles}), plotted on a linear scale to aid comparison. The consistency of 
the counts implies negligible selection bias in either the detection or 
photometry between the shallow and deep fields.}
\end{figure}

In the top panel of Figure 5 we plot in the traditional way, the galaxy
number counts for both the 'field' (as per Figure 4) and what we
measure in the direction of A2218. We also supplement the counts at
bright magnitudes ($F606W\le 17.5$) by plotting those derived from
the ground-based Millennium Galaxy Catalog (MGC) by Liske et al. (2003). 
We caution that this has involved transforming the MGC counts from
the $B$-band (in which they were measured) onto our $F606W$(VEGA)
system. This we did using the relation  
$B_{\rm MGC}- F606W$(VEGA)$=1.06$ derived by Driver et al 2003. 
This simple colour conversion is only strictly valid for the very bright
end of the MGC counts: $F606W < 17.5$\,mag. However, since their
only purpose in this study is to provide a bright magnitude `anchor'
point -- the accuracy of which has a negligible effect on the overall
accuracy of our field galaxy subtraction at the much more critical
fainter magnitudes -- this is quite satisfactory. 

Comparison of the A2218 and field counts in the upper panel of Figure 5 
shows a clear divergence between the two with decreasing magnitude,
with the field counts having a slope close to the canonical value
of $\sim 0.4$ observed in other studies at these wavelengths (e.g., 
Metcalfe et al. 2001). To provide a clearer picture of the 
deviation between the A2218 and field counts, we plot in the lower
panel of Figure 5 the counts relative to the 
$N\propto 10^{0.4\,m}$ relation (where $m$ represents our F606W
magnitudes), appropriately
normalised to the A2218 mosaic field of view. This reveals some
residual magnitude dependence in the field counts, so to account
for this and ensure that they are accurately represented over the
entire magnitude range covered by our study, we fit a quadratic to 
the combined MGC+WFPC2 counts, yielding:
\begin{equation}
{d\log_{10}N\over dm}=-12.9346+0.96077m-0.012874m^2.
\end{equation}
This quadratic fit is shown as the solid line in both panels of 
Figure 5.

\begin{figure}
{\includegraphics[width=5.7cm,angle=270]{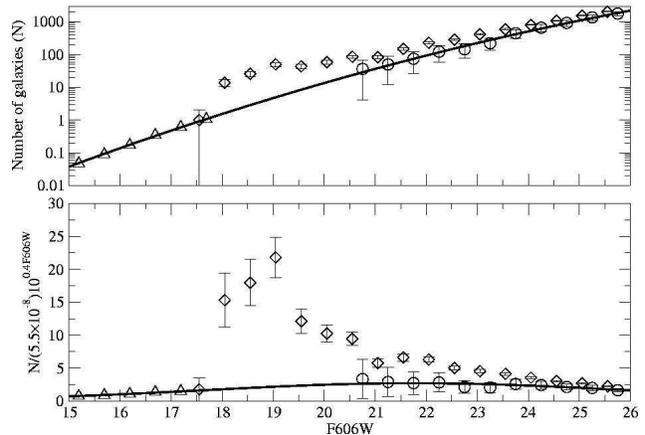}}
\caption{{\it Top panel} Galaxy number counts as a function of F606W
magnitude for Abell 2218 ({\it diamonds}), the combined HDF-North, HDF-South
 and WFPC2 counts ({\it open circles}) and the MGC counts ({\it triangles}). The
{\it solid line} is a polynomial fit to all the reference field counts.
{\it Bottom panel} The galaxy number counts shown relative to the 
$N\propto 10^{0.4\,m}$ relation.}
\end{figure}

Table 1 provides a tabulation of the galaxy counts (and their associated
errors) relevant to our A2218 analysis. In column 2 we list the total 
counts, $n_{tot}$, observed within our A2218 field and in column 5 our estimates of
the numbers of field galaxies, $n_{ref}$, within the A2218 mosaic field-of-view,
based on the above quadratic fit. Columns 3 and 6 give estimates for the Poisson
noise in the components that make up each observed count:
the standard deviation of the counts along the cluster line of sight:
\begin{equation} 
P_{\rm tot}=\sqrt{n_{tot}},
\end{equation}
and the error due to Poisson noise in our estimate of the mean field (or reference) count:
\begin{equation}
P_{\rm ref}=\sqrt{{\Omega_{\rm A2218} \over \Omega_{\rm WFPC2}}{n_{ref} \over 12}}, 
\end{equation}
respectively. Here $\Omega_{\rm A2218}$ and $\Omega_{\rm WFPC2}$ represent
the area of sky covered by our A2218 observations and within one WFPC2 field,
respectively.  

There is an additional uncertainty in the galaxy counts as a result of 
general galaxy clustering along the line of sight to A2218 (Glazebrook et
al. 1994, Djorgovski et al. 1995). The error introduced by this effect 
has been computed by Huang et al. (1997) and is given by their 
equations 5 and 6.  An analogous formula is also given by Driver et al. 
(2003; equation 7). This is a vital component which typically dominates the
error budget at faint magnitudes and has often been neglected in previous
studies. The errors introduced into our counts by general galaxy clustering
along the line of sight to the A2218 field ($C_{\rm c}$) and 
to the reference fields ($C_{\rm ref}$) are given in columns 4 and 7, 
respectively. They have been calculated using the Driver et al. equation:
\begin{equation}
\epsilon^2(N)=({\sqrt{2} \over 3})^{-0.8} \cdot \Omega^{-0.4}\cdot
N^2(m)\cdot{10^{-0.235m+2.73}}.
\end{equation}
Here, $N$ is the field number counts (per 0.5\,mag interval) and $\Omega$ is the
area of sky over which the counts are measured. The field galaxy counts 
appropriate to our A2218 field (and hence to the derivation of the `clustering'
errors listed in column 4) are given in column 5 of Table 1. These have
been derived using the above quadratic representation of the
MGC+WFPC2 counts, with the appropriate normalisation applied
to account for the different sky coverage: $\Omega_{\rm A2218}=0.0228$\,sq. deg, 
$\Omega_{\rm MGC}=30.310$\,sq. deg, $\Omega_{\rm WFPC2}=0.0011$\,sq. deg. 
The small area of the WFPC2 field means that the counts observed within
it are more susceptible to line-of-sight clustering, as can be seen in
the larger `clustering' errors listed for the field in
column 7 of Table 1. 

In Figure 6 we show how the observed variance in the galaxy number counts 
from field to field  compares with what we would expect based on
our above error budget analysis. It can be seen that, overall, there
is good agreement between the two, with our predictions, if anything, 
being slightly pessimistic.

\begin{figure}
{\includegraphics[width=5.7cm,angle=270]{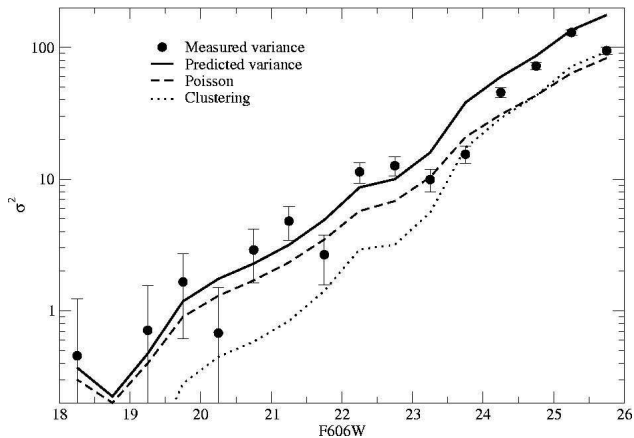}}
\caption{Comparison of the measured variance in the 12 reference fields
({\it filled circles}) and the predicted variance as described in the text
({\it solid line}).  The {\it dashed line} and {\it dotted line} show the
Poisson and clustering contributions to the predicted variance respectively.
Note that the error bars plotted for the observed points were determined
using a `jack-knife' method.}
\end{figure}

The final column of Table 1 contains the field-subtracted
`cluster' counts within our A2218 field, with their quoted uncertainties
representing the combination of all the contributing errors described 
above. Figure 7 shows graphically the contribution of each of the four errors to 
the final error budget.

\vspace*{6pt}
\begin{table*}
\setcounter{table}{0}
\centering
\begin{minipage}{140mm}
\begin{tabular}{|l|l|l|l|l|l|l|l|l|} \hline

$F606W$ &$n_{tot}$&$P_{tot}$& $C_{\rm c}$ &$n_{ref}$&
$P_{\rm ref}$ &  $C_{\rm ref}$ &$n_{clus} (n_{tot}-n_{ref})$ \\ \hline
17.55 & 1 & 1 & 0.53 & 0.92 & 1.23 & 0.27 & $0.1\pm 1.7$ \\
18.05 & 14 & 3.74 & 0.82 & 1.63 & 1.64 & 0.42 & $12.4\pm 4.2$ \\
18.55 & 26 & 5.10 & 1.26 & 2.87 & 2.18 & 0.65 & $23.1\pm 5.7$ \\
19.05 & 50 & 7.07 & 1.91 & 4.97 & 2.86 & 0.98 & $45.0\pm 7.9$ \\
19.55 & 43 & 6.56 & 2.85 & 8.47 & 3.74 & 1.46 & $34.5\pm 8.2$ \\
20.05 & 59 & 7.68 & 4.18 & 14.23 & 4.85 & 2.14 & $44.8\pm 10.2$ \\
20.55 & 86 & 9.27 & 6.05 & 23.57 & 6.24 & 3.10 & $62.4\pm 13.1$ \\
21.05 & 83 & 9.11 & 8.62 & 38.46 & 7.97 & 4.42 & $44.5\pm 15.5$ \\
21.55 & 151 & 12.29 & 12.11 & 61.82 & 10.11 & 6.20 & $89.2\pm 20.9$ \\
22.05 & 229 & 15.13 & 16.75 & 97.92 & 12.72 & 8.58 & $131.1\pm 27.3$ \\
22.55 & 288 & 16.97 & 22.84 & 152.82 & 15.89 & 11.70 & $135.2\pm 34.6$ \\
23.05 & 414 & 20.35 & 30.67 & 234.99 & 19.70 & 15.71 & $179.0\pm 44.6$ \\
23.55 & 594 & 24.37 & 40.59 & 356.01 & 24.25 & 20.79 & $238.0\pm 57.1$ \\
24.05 & 812 & 28.50 & 52.92 & 531.42 & 29.63 & 27.10 & $280.6\pm 72.3$ \\
24.55 & 1089 & 33.00 & 67.99 & 781.60 & 35.93 & 34.82 & $307.4\pm 90.6$ \\
25.05 & 1539 & 39.23 & 86.05 & 1132.64 & 43.25 & 44.07 & $406.4\pm 113.0$ \\
25.55 & 2044 & 45.21 & 107.32 & 1617.20 & 51.68 & 54.97 & $426.8\pm 138.8$ \\ \hline
\end{tabular}
\caption{Summary of the number counts for A2218. The total galaxy counts detected in the A2218 field 
(column 2) and the field reference counts (column 5) along with all the associated errors (see text).
The final column gives the field subtracted galaxy counts along with the combined error.}
\end{minipage}
\vspace{1cm}
\end{table*}
\begin{figure}
{\includegraphics[width=5.7cm,angle=270]{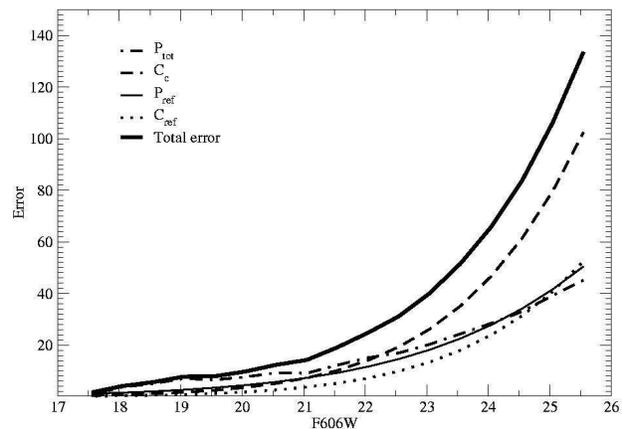}}
\caption{The errors in the Abell 2218 number counts as a
function of magnitude. The four individual error components are shown along
with the combined total error ({\it solid line}).}
\end{figure}

\subsection{Lensing}
A further possible correction to the background counts concerns 
the effects of gravitational lensing of the background population
by the cluster mass (Trentham 1998). This consists of two competing effects:
a magnification effect, where the background population is brightened 
and therefore seen in greater numbers at any given magnitude, and an
apparent expansion of the background field behind the cluster, which
reduces the surface density of background objects. The precise details 
of the combined correction required to account for these two effects
are given in Bernstein et al. (1995) and Trentham (1998); we therefore 
give here only a brief outline of the model which we will
use to estimate the effect of lensing in A2218. 

The quantity that requires computation is the fractional change in the
counts due to lensing:  
\begin{equation}
f_{\rm lens}={N^\prime (m) \over N(m)},
\end{equation}
where $N(m)$ is the observed reference counts per unit area and
$N^\prime (m)$ is the counts we should instead observe when 
the lensing effect of the cluster is taken into account.
Calculation of $N^\prime (m)$ requires knowledge of 
the number of galaxies per unit redshift per magnitude per
steradian, which is given by:
\begin{equation}
n(m,z)={\ln(10) \over 10\pi}\phi (m,z) {dV \over dz},
\end{equation}
where $\phi (m,z)$ is the field galaxy luminosity function at
redshift $z$. We assume that this can be described by a Schechter 
function at all redshifts, with luminosity evolution parameterized 
by:
\begin{equation}
\phi^*(z)=\phi^*(0)(1+z)^2
\end{equation}
\begin{equation}
M^*(z)=M^*(0)+5\log(1+z)
\end{equation}
(see Broadhurst et al. 1995, Trentham 1998). 
We use the local luminosity function measured from the 2dF
Galaxy Redshift Survey (Norberg et al. 2001)
and the colour transformations from Driver et al. (2003) which, after
conversion to our cosmology, results in 
$\phi^*=0.0169{\rm Mpc}^{-3}$, $M^*=-20.65$ and $\alpha=-1.21$.
The volume element is given by: 
\begin{equation}
{dV \over dz}=\Omega x^2 {dx \over dz},
\end{equation}
where $x$ is the comoving distance, and for a flat cosmology:
\begin{equation}
{dx \over dz}={c \over H_{0}\sqrt{\Omega_{m}(1+z)^3+\Omega_{\Lambda}}}.
\end{equation}
Having defined the model $n(m,z)$ distribution,  the true $N(m,z)$ can be
recovered by requiring that ${\int_{0}}^{\infty} N(m,z)dz$ reproduces the
observed number counts, that is: 
\begin{equation} 
N(m,z)={N(m) \over {\int_{0}}^{\infty} n(m,z)dz}n(m,z).
\end{equation}
The function $N^\prime (m)$ is then given by:
\begin{equation}
N^\prime (m,z)={1 \over A}N(m+2.5\log A,z),
\end{equation}
where $A$ is the amplification of the lens. Again, following Trentham,
we assume the cluster mass profile to be a spherically symmetric
isothermal sphere. $A$ is then given by:
\begin{equation}
A(r,z)=\left|{1-{4\pi \sigma^2 D_{c}D_{cs}(z) \over rc^2
D_{s}(z)}}\right|^{-1}
\end{equation}
where $D_{c}$, $D_{s}$ and $D_{cs}$ are the angular diameter 
distances to the cluster, to the source and from the cluster to the source respectively
(Miralda-Escude 1991). $f_{lens}$ can then be calculated by radially averaging 
${N^\prime \over N}$ over the annulus for which we want to 
evaluate the lensing correction:
\begin{equation}
f_{lens}(m)={2\int_{r_{1}}^{r_{2}}\left[\int_{0}^{z_{c}}N(m,z)dz+\int_{z_{c}}^{\infty}
N^{\prime}(m,z,r)dz\right] rdr \over r_{2}^{2}-r_{1}^{2}},
\end{equation}
where $r_{1}$ and $r_{2}$ are the inner and outer radii of the annulus. 

In Figure 8 we plot $f_{\rm lens}$ for a series of annuli in which we 
will derive cluster LFs in our subsequent analysis. The {\it solid} line 
shows $f_{\rm lens}$ for the inner region of the cluster bounded 
by $r_{1}=100$\,kpc and $r_{2}=300$\,kpc. It is this region
where the effects of lensing are at their maximum. Note
that the inner 100\,kpc has been excluded due to the unrealistic nature of
the isothermal model in that region (AbdelSalam, Saha \& Williams 1998). 
The {\it dotted line} shows $f_{\rm lens}$ for the outer region of the cluster,
with boundaries $r_{1}=300{\rm kpc}$ and $r_{2}=750{\rm kpc}$. Finally, the
{\it dashed line} shows  $f_{\rm lens}$ for an annulus encompassing the
`whole' field: $r_{1}=100{\rm kpc}$ to $r_{2}=750{\rm kpc}$.  

In general, the family of curves shown in Figure 8 are in excellent
agreement with those derived by Trentham (for Abell~665 at $z=0.18$; 
see his Figure 3) in terms
of their overall behavior with apparent magnitude, the amplitude of the
effect ($\ls 10\%$), and its much diminished importance at larger
clustercentric radii. 
The only systematic difference of note is that our $f_{\rm lens}$ 
function drops below unity at faint magnitudes ($F606W>23$) -- that is, 
lensing {\it decreases} (rather than boosts) the number of faint
background galaxies seen in the direction of the cluster. This difference 
can be attributed to slightly different
assumptions made about the $n(m,z)$ distribution, in particular the 
differences in the representations of the observed galaxy counts, $N(m)$, 
in normalizing $n(m,z)$. Here Trentham uses a linear function to 
represent $d\log N/dm$, whereas we have used a quadratic representation 
(see Figure 5).

\begin{figure}
{\includegraphics[width=5.7cm,angle=270]{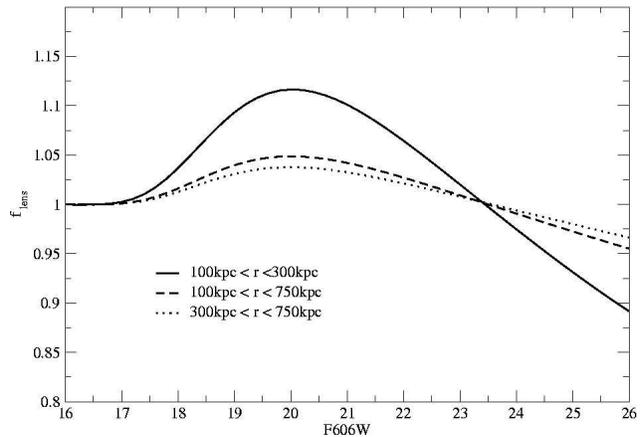}}
\caption{The function $f_{\rm lens}$, averaged over three different
annuli: $100{\rm kpc} < r < 300{\rm kpc}$ ({\it solid line}), 
$100{\rm kpc} < r < 750{\rm kpc}$ ({\it dashed line}) and 
$300{\rm kpc} < r < 750{\rm kpc}$ ({\it dotted line}).   }
\end{figure}

\section{Results}

\subsection{Luminosity Function}
As a first step in our analysis, we take the final field-corrected cluster
counts listed in the final column of Table 1 and construct the
luminosity distribution (LD) for A2218's galaxy population; this
is shown in Figure 9 ({\it filled} circles). 
Formally, this distribution is well described by a Schechter 
function with $M^*_{\rm F606W}=-20.52\pm 0.27$ and 
$\alpha = -1.38 \pm 0.05$ over the range $ -21.5 <M_{\rm F606W} < -13.2 $ 
(solid line in left panel of Figure 9). Because the errors in $M^*$ and
$\alpha$ are tightly correlated, we present alongside our derived LD 
the error contours for these two parameters (the {\it thick} lines
in the right-hand panel). Also shown in this figure is the  
LD recovered after the reference field counts have been scaled by $f_{\rm lens}$ 
({\it dashed} line in Figure 8) to correct for lensing effects ({\it open
circles}). {\it It can be seen that this makes very little difference to
our observed LD.} This is reflected quantitatively in the values of the
fitted Schechter function parameters, $M^*_{\rm F606W}=-20.25\pm 0.28$ and
$\alpha = -1.40 \pm 0.05$, which differ from the original values by
less than $1\sigma$. 

For comparison, we also plot in Figure 9 the $M^*_{\rm F606W}$ and
$\alpha$ values for the Schechter function fits to the composite
LF derived for the $z\sim 0.1$ clusters in the 2dF Galaxy Redshift 
Survey (2dFGRS; De Propris et al. 2003), and the LF derived for
the rich cluster Abell~868 ($z=0.15$) from a comparable 
HST-based dataset to what we have used here, by Driver et al. 
(2003). It can be seen that our A2218 LF has a steeper faint-end
slope (more negative value of $\alpha$) and a slightly fainter
$M^*$ than these other two LFs, particularly for the 2dFGRS 
clusters where, given the size of the error bars, the differences are
reasonably significant. On the other hand, the 2dFGRS and A868 LFs 
do not extend as faint as ours ($M_{\rm F606W}\simeq -16$ cf.
$M_{\rm F606W}\simeq -13$), and it is well known that the value 
of the parameter $\alpha$ is quite sensitive to the luminosity range
covered, with the faint end steeping at fainter magnitudes (Lobo et al.
1997). A better comparison is obtained by fitting the A2218 data over the
same absolute magnitude range, for which we recover $M^*_{\rm F606W}=
-20.26 \pm 0.27$ and $\alpha =-1.23 \pm 0.09$. This is also the range over
which our sample is complete with respect to surface brightness.
The error contours for the LF parameters fit over this range are shown as the {\it thin}
contours in Figure 9. We see that the LF in this case has a shallower slope, 
with its $\alpha$ value being consistent with those of the 2dFGRS and
A868 LFs. However, the small offset in $M^*$ still remains. Such small
differences may well result from the regions of the clusters being
compared here not being identical; the likelihood of this being the
case is highlighted below. 
 
Finally, it is also worth noting that we see evidence of an `inflexion'
in A2218's LF at $M_{\rm F606W}\approx-18$, something which has been
seen in many other clusters (e.g. Driver et al. 1998a; Barkhouse,
Yee \& Lopez-Cruz 2002).

\begin{figure}
\vspace{0.3cm}
{\includegraphics[width=4.5cm,angle=270]{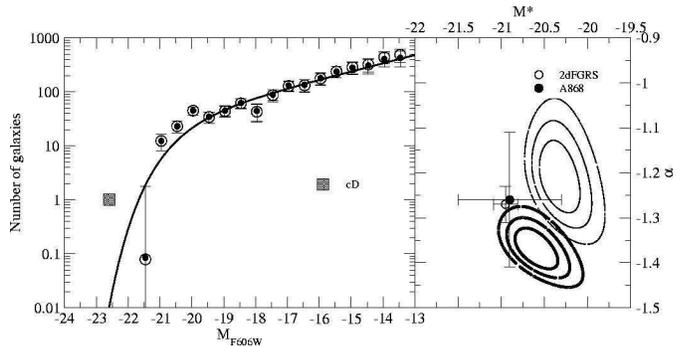}}
\caption{{\it Left panel:} The galaxy luminosity distribution derived
for A2218. The data are represented by {\it circles}; those
that are {\it filled} have had no correction for the effects of
gravitational lensing, whereas the {\it open circles} include this
correction. The data point contributed by the central cD galaxy
is shown as the {\it filled square}.
A Schechter function fit to the (uncorrected) data points
is represented by the {\it solid line}. {\it Right panel:} The 
$1\sigma$, $2\sigma$ and $3\sigma$ error contours for the Schechter 
function parameters $M^*$ and $\alpha$ for $M<-13.2$ 
({\it thick contours}) and $M<-16$ ({\it thin contours}) . For comparison,
the values of these parameters for the 2dFGRS composite cluster LF (De
Propris et al. 2003) and that of A868 (Driver et al. 2003) are represented
by the {\it open} and {\it filled} circles, respectively.}
\end{figure}

The wide field provided by the 22 WFPC2 pointings allows us to
sub-divide the cluster into projected `core' and `halo' components and
re-derive the luminosity distributions for each component.  To do this
we split the inner and outer regions at $300 h^{-1}$\,kpc from the
central, dominant cD galaxy.  Somewhat arbitrarily, this radius was chosen
to be twice the typical cluster core radius, as measured, for example, for
the ENACS sample (Adami et al. 1998). In assembling the luminosity
distribution for the inner region, we exclude the central $100 h^{-1}$ kpc
because of our inability to compute lensing corrections in this region
(see Section 2.3). The outer radial limit is $\approx 750 h^{-1}$ kpc. 

Figure 10 shows the resulting LDs with both the uncorrected ({\it solid
symbols}) and lensing corrected ({\it open symbols}) data points
plotted. Our Schechter function fits and error-contours are shown in the
usual way; because the lensing corrections have a negligible effect, 
only those based on the uncorrected data are displayed. From Figure 10
we see that the LD for the central region has a significantly flatter
faint-end slope than that for the outer region. The inner `core' LF has 
a slope of $\alpha=-1.23 \pm 0.13$ while the outer `halo' LF has
$\alpha=-1.49 \pm 0.06$. This is in line with recent claims
that the LF of galaxies in the Coma cluster is flatter in the vicinity
of the two central dominant giants, with it steepening as a function of
clustercentric distance (Beijersbergen et al. 2002; see also Lobo et al. 
1997). Driver et al. (2003) obtained a a similar result from their
LF analysis of A868. Barkhouse et al. (2002) also find a significant
steepening of the LF as a function of clustercentric radius in their sample
of local clusters as do Sabatini et al. (2003) for the Virgo cluster and
Andreon (2002) for the $z=0.31$ cluster A2744.  Similarly, luminosity segregation
may be responsible for the dwarf population density relation discussed in the 
following two sections. However, Mobasher et al. (2003) argue against the claim of
Beijersbergen et al. (2002) for luminosity segregation in Coma, while Paolillo 
et al. (2001) fail to find steeper LFs in their composite LF for the outer regions
of nearby clusters.

\begin{figure}
\vspace{0.3cm}
{\includegraphics[width=4.5cm,angle=270]{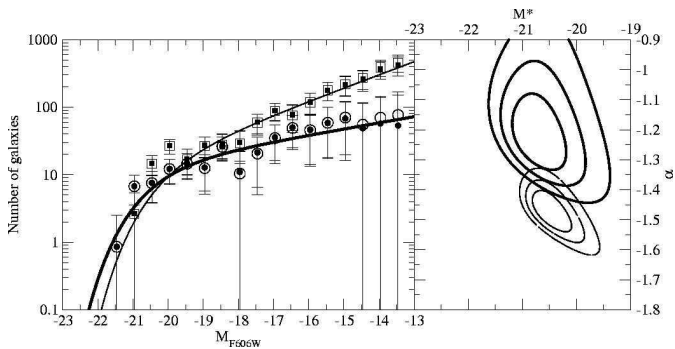}}
\caption{{\it Left panel:} Galaxy luminosity distributions (LDs) 
for A2218, split on the basis of clustercentric radius. The 
LD for galaxies in the inner `core' region of the cluster 
($100 \leq r < 300$\,kpc) is represented by {\it circles}; those
that are {\it filled} have had no correction for the effects of
gravitational lensing, whereas the {\it open circles} include this
correction. The LD for galaxies in the outer `halo' region of the cluster 
($300 \leq r < 750$\,kpc) is represented by {\it squares}, differentiated
in the same way to show the effects of lensing corrections. 
Shechter function fits to the uncorrected data are shown as the {\it thick}
and {\it thin} solid lines for the inner and outer regions, respectively.
{\it Right panel:} The $1\sigma$, $2\sigma$ and $3\sigma$ error ellipses for
the Schechter function parameters, $M^{*}_{\rm F606W}$ ({\it horizontal}
axis) and $\alpha$ ({\it vertical} axis), derived from the fits shown
in the left-hand panel; inner region ({\it thick lines}), outer
region ({\it thin lines}).}
\end{figure}

\subsection{Dwarf-to-giant ratio}

A simpler, non-parametric method to test for the luminosity segregation 
indicated in Figure 10, is to measure the
dwarf-to-giant ratio (DGR -- Driver et al. 1998a). This
is much less sensitive to the correlated errors that affect the 
Schechter LF parameters.  To do so, we define `giants' as galaxies brighter
than $M_{\rm F606W}=-18$ and `dwarfs' as galaxies with $-18 < M_{\rm F606W}
<-15$. The DGR -- the ratio of the {\it numbers} of
so defined dwarfs to giants -- was calculated for each dwarf galaxy by
finding the distance to its 10th nearest giant galaxy and counting the
number of dwarf galaxies within the same region. The area of the circle was
calculated numerically to take into account gaps in the mosaic (Figure 1).
The local galaxy density (Dressler 1980) was then simply taken as 10 
divided by this area.
In order to reduce errors, the data were then binned according to the local
giant galaxy density measured in the vicinity of each dwarf
galaxy. Field subtraction was performed using the smoothed counts from
Figure 5. 

Figure 11 shows the DGR plotted as a function of local giant galaxy
density. Our results show clearly that the DGR in A2218 decreases 
smoothly and monotonically with increasing local giant 
galaxy density, consistent with the dwarf population--density relation 
proposed by Phillipps et al. (1998). Indeed our data define the
relation much more cleanly and over a large range in galaxy density
than the data from Smith et al. (1997) and Driver et al. (1998a) 
(which we also plot in Figure 11 for a qualitative comparison) 
upon which Phillipps et al's proposed relation was based. 
That these latter data points all appear to sit lower on the diagram 
with respect to our A2218 data
is most likely attributable to differences in filters, magnitude
depths and binning in galaxy density between the different studies.

Our data are of sufficient depth to allow us to also consider a fainter
sample of `ultra-dwarf' galaxies, with $-13.5 < M_{\rm F606W} < -15$. The
DGR values for this population are also plotted in Figure 11 and appear to
follow the same trend with density as the brighter sample, indicating
commonality between the ultra-dwarf and dwarf populations.

\begin{figure}
{\includegraphics[width=5.7cm,angle=270]{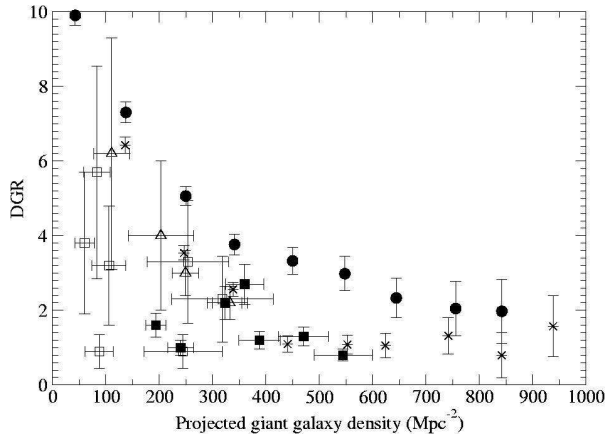}}
\caption{The variation of the dwarf-to-giant galaxy ratio (DGR) with local 
giant galaxy density. {\it Filled circles} represent the `intermediate' 
dwarfs ($-18<M_{\rm F606W}<-15$) in A2218; {\it starred points} represent 
the `ultra-dwarfs' ($-15<M_{\rm F606W}<-13.5$) in A2218. The error bars
include Poisson contributions only and lensing effects are ignored. The 
{\it open triangles} show the data for A2554 from Smith et al. (1997). 
The {\it filled squares} represent the central regions of the clusters from
Driver et al.  (1998a) and the {\it open squares} the outer regions.}
\end{figure}

\subsection{The radial profile of galaxies in A2218}

In order to determine whether or not the decreasing DGR is due only to
the increasing giant galaxy density toward the cluster core or whether there
is a corresponding decrease in the number of dwarfs for higher giant
galaxy densities, we have derived radial density profiles for these
different sub-populations within A2218. These are plotted in Figure 12,
with the data for the giants ($M_{\rm F606W}<-18$) shown in the 
{\it top panel}, those for the `intermediate' dwarfs ($-18<M_{\rm F606W}<-15$)
shown in the {\it middle panel}, and those for the `ultra-dwarfs' 
($-15<M_{\rm F606W}<-13.5$) shown in the {\it bottom panel}. Again
the data with and without corrections for lensing are plotted, and it
can be that the difference between them is negligible. Accordingly, in
fitting the data with a power-law ($\sigma(r) \propto r^{a}$; 
{\it solid lines} in Fig. 12), we do so to the uncorrected data
only. Note that we also exclude the innermost radial point
in these fits as well (corresponding to the inner $100h^{-1}$ kpc
where the lensing correction is uncertain.

The giant galaxy projected density is seen to drop off with a power-law
index $a=-1.13\pm0.15$, consistent with that of a projected isothermal
sphere ($\propto a^{-1}$).  We also note that beyond a radius of 
$\sim 1$\,arcmin ($\sim 128$\,kpc), it traces very well the projected
X-ray gas density profile ({\it dashed} line) observed for A2218 (Machacek
et al. 2002), which has a core radius of $r_{\rm c}=66.4^{\prime\prime}$ and 
$\beta=0.705$. 

For the intermediate luminosity dwarfs, the distribution is flatter than that
of the giants and is best fit by a power-law with index
$a=-0.63\pm0.09$. In contrast, the ultra-dwarf galaxy surface density shows
a deficiency in the centre of the cluster, and then is approximately flat
at $r>0.6$\,arcmin ($\sim 200\,h^{-1}$\,kpc). A power-law fit to the data
has a index of $a=-0.03\pm 0.13$. The most likely explanation for this 
behavior is that the fainter galaxies have a more
extended distribution, and hence larger scale radius, than the giants, 
as also seen for the bright and faint dEs in the Coma cluster 
(Secker et al. 1997).

\begin{figure}
{\includegraphics[width=8cm,angle=270]{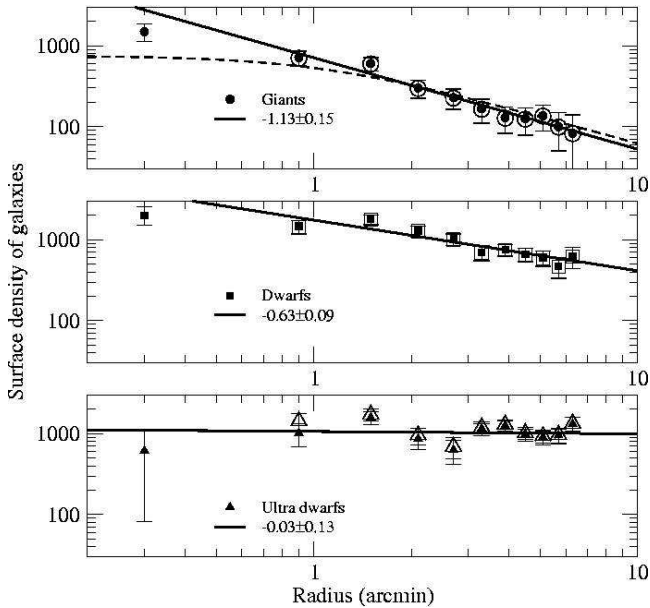}}
\caption{Radial surface density distributions for the giant, dwarf, 
and ultra-dwarf galaxy populations in A2218: {\it Top panel:} the giant
($M_{\rm F606W}<-18$) galaxies, with the uncorrected data shown as {\it
filled circles}, and the lensing-corrected data shown as {\it open circles}.
The projected X-ray gas density profile is shown as the {\it dashed line}. 
{\it Middle panel:} the intermediate dwarf ($-18<M_{\rm F606W}<-15$)
galaxies, with the uncorrected data shown as {\it filled squares} and 
the lensing-corrected data shown as {\it open squares}. {\it Bottom panel:} 
the ultra-dwarf ($-15<M_{\rm F606W}<-13.5$) galaxies, with the uncorrected data
shown as {\it filled triangles} and the lensing-corrected data shown as {\it open
triangles}. A power-law fit to the uncorrected data points is represented in
all three panels by the {\it solid line}.}
\end{figure}

There are two competing effects which should be considered in interpreting
Figure 12. Firstly, because of the larger galaxy density in the centre of the 
cluster there is a diminishing area effect where the field of view in which 
dwarf galaxies can be detected is reduced due to the presence of brighter 
galaxies (see Driver et al. 1998b). To determine the extent of this effect,
we sum together the areas assigned to detected objects by {\sc SExtractor} 
in both the very central region of the cluster and in the lowest density
regions.  We find in the central region the  total area associated with
objects by {\sc SExtractor} is 8.2\% of the total area, while in the
outskirts only 4.2\% of the area is occupied by objects. Hence
if the data in Figure 12 were to be corrected for this effect, it would amount
to only a 4\% differential adjustment between the innermost (highest density)
and outermost (lowest density) bins. This would have a negligible effect 
in terms of the overall trends seen in the bottom two panels.    

Secondly, there is a radial dependence on the volume of cluster sampled per unit 
area projected on the sky. Assuming the cluster fills a sphere of radius $R$, 
then the volume per unit area projected onto a circle of radius $a$ centered on 
the cluster is :
\begin{equation}
{{\rm Volume} \over {\rm Area}}\approx 
{4 \over 3a^{2}}(R^{3}-(R^{2}-a^{2})^{3 \over 2}).
\end{equation}
Hence the volume per unit area decreases with radius. If we assume a
physical radius for A2218, then we can calculate the corrections in number
needed in each of the radial bins in Figure 12. For a spherical cluster
with a $1.5\,h^{-1}$\,Mpc radius (approximately the Abell radius), a
factor of 1.15 increase in number is required in the outermost annuli 
(with respect to the innermost annuli) to convert from number per
unit area to number per unit volume.  This more than compensates for
the correction (in the opposite sense) due to the diminishing area
effect. So it is likely that the differences in profile shape 
for the giant, dwarf, and ultra-dwarf galaxy populations 
are as large, if not larger than that seen in
Figure 12, and the flattening of profile shape with decreasing
luminosity is a real effect, when both these effects are taken into account.

\section{Summary and discussion}

We have exploited deep, wide-field HST imagery of the well known
rich cluster A2218 at $z=0.18$, to conduct a census of its galaxy 
population over a $\sim 10$\,magnitude range in luminosity ($-23 < M_{\rm 
F606W} < -13$), and thereby explore how its constituent `giant' and 
`dwarf' galaxy populations vary in number as a function of location and 
environment within the cluster. 

Taking the complete ensemble of A2218 galaxies that are identified, statistically,  
within the entire central 1.34\,Mpc$^2$ $h^{-2}$ field studied
and over the full range in luminosity and surface brightness 
($\mu_{F606W}\ls 24.7$\,mag\,arcsec$^{-2}$) probed, we find
their luminosity distribution to be well fitted by a Schechter function,
with parameters $M_{\rm F606W}^*=-20.52\pm 0.27$ and $\alpha=-1.38\pm 0.05$. 
In terms of the faint-end slope, at least, this is in reasonable agreement with 
the LFs measured for other rich clusters, both at similar and more
nearby redshifts. 

Importantly, however, our study has also revealed 
that such `global' measures of the LF represent the superposition of 
a number of important underlying dependencies on clustercentric radius 
and, more fundamentally, the local galaxy density.
The most conspicuous of these is a change in the {\it shape} of A2218's 
LF with clustercentric radius, in particular its faint end having a
steeper slope at larger radii. This indicates that the relative balance
between the numbers of dwarf and giant galaxies varies with environment in 
the cluster, and we have quantified this by measuring the ratio of dwarf 
to giant galaxies (DGR) as a function of local projected galaxy density. 
We find the DGR in A2218 to decrease monotonically with increasing local 
density, dropping by an order of magnitude (from $\sim 10$ to $\sim 1$)
in going from the lowest ($\sim 100$\,gals\,Mpc$^{-2}$) to the highest 
($\sim 1000$\,gals\,Mpc$^{-2}$) densities probed by our observations. 
This confirms the existence of a ``dwarf galaxy--density relation'' as 
first suggested by Phillipps et al. (1998). Furthermore, we have shown 
that this relation arises from galaxies of different luminosity having 
quite different radial surface density profiles. The brightest, giant 
galaxies have the steepest profile, consistent with an isothermal sphere 
and that of the hot X-ray gas. The fainter, dwarf galaxies have much 
shallower profiles, indicating that they are a much more extended 
population with a much larger scale radius. Indeed the `ultra-dwarf' 
galaxy population in A2218 is found to have a radial profile which is 
essentially flat, indicating a constant surface density at least over 
the central 750\,kpc studied here.  

This quite clear `segregation' between galaxies in these different
luminosity regimes could be a result of primordial conditions, evolution, 
or both. As has been previously discussed by Phillipps et al. (1998), 
the numerical CDM models of Kauffmann et al. (1997) point to the dwarf 
population--density relation being a natural consequence of the 
hierarchical clustering process. In this picture the dwarf galaxies, 
which form via the gravitational collapse of small amplitude density peaks 
in the primordial universe, are predicted to be more numerous and less 
clustered than their brighter counterparts.

However, Moore et al. (1998) have also been able to explain the 
dwarf population--density relation as the result of galaxy `harassment', 
which operates more effectively in denser regions. As the cluster tidal 
field becomes increasingly important in the cluster centre, the smaller 
spheroidal galaxies will be destroyed in the innermost part of the cluster 
and global tides will disintegrate the lowest surface brightness objects 
into the diffuse stellar background. In the Coma cluster, for example, 
Biviano et al. (1996) find that the cluster structure is better traced by 
the faint galaxy population, which forms a single smooth structure,  with 
the brighter galaxies being located in sub-clusters. They interpreted this 
as evidence for the ongoing accretion of groups onto the main body of the 
cluster. In this context, it is interesting to note that Bernstein et al. 
(1995) find a significant flattening of the density profile of faint 
galaxies in the inner 100\,kpc of the Coma cluster.

The deficiency of dwarf galaxies in the centre of Abell 2218 may also be related 
to the presence of a cD galaxy. cD galaxies probably form suddenly during the 
collapse and virialization of compact groups or poor clusters which then merge 
with other groups to form a rich cluster (Merritt 1985). Lopez-Cruz et al.
(1997) suggest that the flatter faint-end slope of the galaxy LF which they
find in rich clusters, in particular clusters that contain cD galaxies, 
results from the disruption of large number of dwarf galaxies early on in the
clusters evolution. A similar effect is claimed by Barkhouse et al. (2002).
The stars from the disrupted galaxies are redistributed throughout the
cluster, generating the cD halo.  This is consistent with a lack of dwarf
galaxies in the core of Abell 2218.

Ultimately, a single cluster is not sufficient to both establish the trends
we discussed and analyse their dependence on cluster properties. A larger
sample of clusters with deep, wide-field imaging is needed to make progress
on these important issues. This is a project we are currently embarking upon, 
the results from which will be reported in future papers.

\def\ref{\par\noindent\hangindent\parindent}

\section*{Acknowledgments}
M.B.P., W.J.C., and P.E.J.N. acknowledge the financial support of the Australian
Research Council throughout the course of this work. This research has make use
of the NASA/IPAC Extragalactic Database (NED) which is operated by the jet
propulsion laboratory, California Institute of Technology, under contract with
the National Aeronautics and Space Administration.

\clearpage

\label{lastpage}

\end{document}